\documentclass{article}
\newcommand{\be}{\begin{equation}}
\newcommand{\ee}{\end{equation}}
\begin{document}

\baselineskip18pt

\title{Parametric  Dynamics of Level Spacings in  Quantum Chaos}

\author{Piotr Garbaczewski\thanks{Email: pgar@proton.if.wsp.zgora.pl}\\
Institute of  Physics, Pedagogical University \\ PL-65 069 Zielona
G\'{o}ra,  Poland \\}
\maketitle

\begin{abstract}
 We identify    parametric  (radial) Bessel-Ornstein-Uhlenbeck  stochastic
 processes as primitive dynamical  models of energy level repulsion  in irregular
 quantum systems.  Familiar
 GOE, GUE, GSE and non-Hermitian Ginibre   universality classes
 of  spacing distributions  arise as  special cases in that formalism.
 \end{abstract}
PACS Numbers: $02.50, 03.65,  05.45$\\
 \vskip1.0cm

In case of complex   quantum systems, affinities with the
statistical theory of   random-matrix spectral problems
   support the viewpoint \cite{bohigas,haake},  that certain features of a
  fully developed classical chaos  can be elevated to  the quantum level as
universality classes of spectral fluctuations.
However, in those cases, an \it ensemble \rm of large (size is to
grow to infinity) random matrices   is usually set in
correspondence with  \it one \rm  quantum system.

 Then, an immediate question arises:  how may one justify a  comparison of
 a statistical ensemble of randomly disordered spectral series   with
   \it  one \rm only specific  (and generally being or "looking"  random)
    energy level sequence  of an a priori chosen  quantum system.
If specialized to  the quantum chaos context, the basic hypothesis
behind previous arguments is that  quantum
  Hamiltonian may be represented by
just one  random-matrix representative   drawn from an ensemble of suitable
random ones, provided  things happen in the large matrix size regime (needed to
conform with the semiclassical regime in quantum theory).
That  particular  issue  of an \it individual \rm  versus \it
ensemble  \rm spectral information
  is the major objective of  our investigation.

To our knowledge this immediate conceptual obstacle, except for
preliminary investigations of Ref. \cite{valz},  has not  been
seriously addressed in the quantum chaos literature. A  partial
answer to that  question, \cite{haake,pandey}, points towards certain \it
ergodicity \rm properties appropriate for   models of the
parametric level dynamics (Coulomb gas, plasma or else), evolving
in "fictitious time" and ultimately approaching suitable equilibria,
characterized by invariant  probability  measures.

Let us mention that the possibly troublesome  "fictitious" time parameter
 may be consistently interpreted as the running coupling constant measuring the
 strength of the chaotizing perturbation, or  more generally as a
 "complexity parameter" whose growth
 to infinity gives account of  a complexity increase in the quantum system,
 \cite{shukla}.

 That  amounts to  a reinterpretation of random-matrix theory in terms of an
 equilibrium statistical mechanics for a  fictitious $n$-particle  system, where
$n$ needs  to grow indefinitely to meet  (or satisfactorily approximate)
spectral  predictions  for  semiclassically analyzed irregular quantum systems.

A common mathematical  structure behind this argument is related to
 the so-called Calogero-Moser Hamiltonian system, \cite{haake,shukla},
 for $n$ particles  on a real line
(actually energy levels, here $n$ also corresponds to the matrix size in the
corresponding random-matrix theory reasoning and we need  $n$ to be large)
which interact via pairwise  inverse square interaction and a harmonic  attraction.
In particular, the probability distribution of n-particle (energy level) coordinates
in  the ground state of the Calogero-Moser Hamiltonian is known to coincide with the
Wigner-Dyson distribution of energy levels, regarded as  a statistical state of equilibrium
in the  large "fictitious time" asymptotic of the corresponding random-matrix
dynamical model, cf. \cite{haake,shukla}.

It is the \it level repulsion  \rm which is routinely  interpreted  as a
quantum manifestation of classical chaos.
 Normally that is quantified by means of polynomial  modifications  of the
 Gaussian probability law (in association with the Wigner-Dyson  statistics  of adjacent
 level spacings for  e.g. unitary, orthogonal and   symplectic  random matrix ensembles).

For completness, let us list the standard  level spacing formulas:
$P_1(s) =    s {{\pi }\over 2} exp(-{{s^2\pi }\over 4})$,
$P_2(s) = s^2 {{32}\over \pi ^2} exp(-{{s^2 \pi }\over 4})$ and  $P_4(s) =
s^4 {{2^{18}}\over {3^6 \pi ^3}}\, exp(- {{s^2 64}\over {9\pi }})$,
corresponding respectively to the GOE, GUE
and GSE random-matrix theory predictions. The cubic repulsion case
$P_3(s) = s^3 {{3^4\pi ^2}\over  {2^7}}  \exp(-s^2 {{3^2\pi }\over {2^4}})$ is
related to the  non-Hermitian Ginibre ensemble, \cite{haake} and was not covered by
the alternative technique for complex
 spectra analysis, as developed in Ref. \cite{shukla}.

Once  we have encountered probability densities on the positive  half-line in $R^1$, it is
rather natural to investigate a general issue  of   parametric stochastic processes
which would provide  a dynamical model  of   level repulsion in an irregular quantum system
and generate at the same time   spacing
densities as those of  asymptotic invariant (equilibrium)  probability measures.
Such  random processes clearly must  run with respect to the previously mentioned
"fictitious" time-parameter and take values in the set of all  level spacings  which are
appropriate for  a complex quantum system or the corresponding random-matrix ensemble.

Effectively, we wish to introduce a  Markovian diffusion-type  process which might
stand  for a reliable approximation of  a random walk over \it  level spacing sizes.\rm

For future reference
let us mention that in the regime of equilibrium (when an invariant measure appears
 in the large "time"  asymptotic), a sample path of such random walk would take the form of
 an ordered sequence of spacings which are sampled (drawn) according to the prescribed
  invariant probability distribution.
 That is precisely  \it one  \rm explicit example of the  ladder  of energy levels,
 understood as a random sample drawn from a suitable ensemble.

 An analysis of statistical features
of this  spectral sequence  involves an ergodicity notion to stay in   conformity
with the ensemble  evaluation of various averages (carried out with respect to the
invariant density), \cite{lasota,lefever}.

We shall consider the previously listed  GOE, GUE and GSE probability
densities on $R^+$  (up to  suitable rescalings !) as,
 distorted in view of the spacing size normalization,
asymptotic invariant densities  of certain parametric Markovian stochastic
processes whose uniqueness status can be unambiguously  settled.

Let is begin from the observation that probability densities  on $R^+$, of the
characteristic form
$f(x) \sim x exp(-{x^2\over 4})$, $g(x) \sim  x^2 exp(-{x^2\over 2})$ and
 $h(x)\sim {x^4\over 4} exp(-x^2)$
appear notoriously in various quantum mechanical contexts (harmonic oscillator or
centrifugal-harmonic  eigenvalue problems), cf. \cite{calogero,zambrini,blanch}.
Notwithstanding, as notoriously they can be identified in connection with special classes
of stationary Markovian diffusion processes on $R^+$, \cite{karlin}.

Anticipating further discussion, let us consider a Fokker-Planck equation on the positive
half-line in  the form:
\be
\partial _t\rho  = {1\over 2} \triangle  \rho   - \nabla [{{\beta }
\over {2x}} - x)\rho ]
\ee
which may be set in correspondence with the stochastic differential equation
\be
dX_t = ({{\beta }\over {2X_t}} - X_t)dt + dW_t \,
\ee
formally valid for a random variable   $X_t$  with values contained in $(0,\infty )$.
Here $\beta \geq 0$ and $W_t$ represents the Wiener process.

Accordingly,  if $\rho _0(x)$ with $x\in R^+$  is regarded as  the density of distribution
of $X_0$ then for each $t>0$ the function $\rho (x,t)$, solving Eq. (1), is the density of
$X_t$.
In view of a singularity of the forward drift at the origin, we refrain from
looking for strong solutions of the stochastic differential equation (2) and
confine attention to weak solutions only and the associated tractable parabolic problem
(1) with suitable boundary data, cf. \cite{karlin}.

In all those cases a mechanism of repulsion
is modeled by the $1\over x$  term in the forward drift
expression. The compensating harmonic attraction  which is    modeled by
the $-x$ term, saturates the long distance effects of
repulsion-induced scattering   and ultimately yields asymptotic
steady (stationary) probability densities.

To interpret a  density   $\rho (x)$ as an asymptotic
(invariant) density of a well defined  Markovian diffusion process  we shall
utilize the rudiments of  so-called Schr\"{o}dinger boundary
and stochastic interpolation problem,
 \cite{zambrini,blanch,garb}.

 Let us notice that both in case of the standard Ornstein-Uhlenbeck process
 and its Bessel (radial)  variant, we have emphasized the role of
 a stochastic process with an asymptotic invariant density.
 To deduce such processes,  in principle we can   start from
  an invariant density and  address an easier issue of the
    associated measure preserving stochastic dynamics   and next  consider whether
   the obtained    process  would drive  a  given initial density towards
   a prescribed invariant measure.  That feature involves the notion of  exactness of the
   related stochastic process, whose straightforward consequence are the properties of
    mixing and ergodicity of the involved  random dynamics, \cite{lasota}.

  There is a general formula \cite{zambrini,garb} relating the forward drift of
  the  sought for stationary process
  with an explicit functional form of an invariant probability density.
  We confine our attention to
  Markov diffusion processes with a constant diffusion coefficient,  denoted
  $D >0$. Then, the pertinent  formula reads:
  \be
  b(x) =  2D {{\nabla \rho ^{1/2}}\over {\rho ^{1/2}}}\, .
  \ee

 In particular,  for the familiar Ornstein-Uhlenbeck process we have  $\rho ^{1/2} (x) =
({1\over {\pi  }})^{1/4} exp(-{x^2\over 2})$ and  $D= {1\over 2}$, so
 we  clearly arrive at  $b(x)= - x $ as should be.
Quite analogously,  in case of  the GUE-type  spacing density, we have
$\delta ={1\over 2}$ and $\rho ^{1/2}(x) = {2\over {\pi ^{1/4}}} x \exp(-{x^2\over 2})$.
Thus, accordingly $b(x)= {1\over x} - x$.

The very same strategy allows us to  identify a forward drift of the Markovian
diffusion process supported by the GOE-type spacing  density.
By  employing   $\rho ^{1/2}(x)=
\sqrt {2x} exp(- {x^2\over 2})$ and setting $D = {1\over 2}$ we  arrive at the formula:
$b(x,t)= {1 \over {2x}} - x $.

We immediately identify the above forward drifts with the ones appropriate for  the time
homogeneous  radial  Ornstein-Uhlenbeck processes, with a corresponding family of
($N>1$ and  otherwise
arbitrary integer) transition probability densities, \cite{karlin}:
\be
p_t(y,x) = p(y,0,x,t) =
2 x^{N-1} \exp(-x^2)\cdot
\ee
$$
{1\over {1-\exp(-2t)}}  \exp[-{{(x^2 + y^2)\exp(-2t)}\over {1-\exp(-2t)}}]\cdot
$$
$$
[xy\exp(-t)]^{-\alpha }
I_{\alpha }({{2xy\exp(-t)}\over {1-\exp(-2t)}})
$$
where $\alpha = {{N-2}\over 2}$ and $I_{\alpha }(z)$ is a modified Bessel function
of order $\alpha $:
\be
I_{\alpha } (z) = \sum_{k=0}^{\infty } {{(z/2)^{2k+\alpha }}
\over { (k!) \Gamma (k+\alpha + 1)}}
 \ee
while the Euler gamma function has a standard form $\Gamma (x) =
\int_0^{\infty } \exp(-t) t^{x-1} dt$.   We remember that $\Gamma (n+1) = n!$ and
$\Gamma (1/2) = \sqrt {\pi }$.

 The  resultant forward drift has the general form:
\be
b(x) = {{N - 1}\over {2x}} - x\,
\ee
and corresponds to $\beta = N-1$ in the notation of Eqs. (1), (2).

By setting $N=2$ in Eq. (4), and then  employing the series representation
of $I_0(z)$, we easily  recover the
asymptotic invariant density for the process:
$lim_{t\rightarrow \infty } p(y,0,x,t) = 2x\exp(-x^2)$.

We can also analyze the large time asymptotic of $p(y,0,x,t)$, Eq. (4) in case of $N=3$
 which gives rise to  an  invariant  density in the form:
${4\over {\sqrt{\pi }}} x^2 exp(-x^2)$.
That obviously corresponds to the GUE-type case with $b(x)= {1\over x} - x$.

When passing to the GSE case, we are interested in the Markovian diffusion
process which is supported by an invariant probability density
$\rho (x)= {2\over {\Gamma (3/2)}} x^4 \exp(-{x^2})$.
Let us evaluate the forward drift of the
sought for process  in accordance with the recipe (3)(we set $\delta = {1\over 2}$):
$b(x,t) = {2\over x} - x$.

A comparison with the definition (6) shows that we  deal  with
a radial  Ornstein-Uhlenbeck
process corresponding to $N=5$. Accordingly,
the transition probability density of the process  displays an
expected asymptotic: $lim_{t\rightarrow \infty } p(y,0,x,t) =
{4\over {\sqrt{\pi }}} x^4 \exp(-x^2)$.
Here we have exploited
$\Gamma (1/2) = \sqrt{\pi }$ to evaluate $\Gamma (3/2)= {1\over 2}\sqrt{\pi }$.

The formulas (4) and (6)  allow us  to formulate  a hypothesis that novel
universality classes may possibly  be  appropriate for quantifying
quantum chaos.
Straightforwardly,  one can verify that   transition probability densities (4)
refer to asymptotic  invariant densities of the  form:
 \be \rho (x) =  {2\over {\Gamma (N/2)}} x^{N-1} exp(-x^2)
 \, . \ee

   In particular we get a direct  evidence in favor of   $N = 4$,
i. e. $b(x) = {3\over {2x}} - x $,  universality class which in fact corresponds to
the Ginibre ensemble of of  non-Hermitian random matrices, \cite{haake}, where
 a cubic level repulsion appears: $\rho (x)=2 x^3 \exp(- x^2)$.

The formulas (6), (7)  allow us  to expect that  processes corresponding to any  $N > 5$
may  be realizable as well, and thus the related higher-power level repulsion  might have
relevance  in the realm of quantum chaos.

In all considered cases, an asymptotic invariance of probability measures (densities)
 is sufficient to yield  ergodic behaviour.
 For each value of $N>1$ we deal
with an independent repulsion mechanism, albeit all of them belong to the radial
Ornstein-Uhlenbeck family.

We have thus  identified a universal stochastic law (in fact, a
family of the like) behind the functional form of basic,  random-matrix theory
inspired and  named generic, spacing probability densities appropriate for  quantum
chaos.

 A common
feature of those parametric processes is an asymptotic balance
between the radial (Bessel-type) repulsion and the harmonic
confinement (attraction), as manifested  in the general form of forward drifts
$b(x) = {{N-1}\over {2x}} - x$ with $N\geq 1$. Here
 $N = 2,3,5$ correspond respectively to the familiar  GOE, GUE and GSE cases while
 $N=4$ to the cubic level repulsion associated  with the non-Hermitian
 Ginibre ensemble.

Let us emphasize at this point that one should keep in mind a number of  possible
 reservations coming from the fact that neither of "universal"  or "generic"
 laws  can be regarded as a faithful representation  of a real state of affairs.
 Usually exact laws are derived for two by two (hence of the small size !)
 random  matrices, and are known to  reappear again
 as approximate spacing formulas in the large random-matrix size regime. That in
 turn allows to achieve a correspondence with semiclassical quantum spectra
 of complex systems.

At the moment we cannot propose a definitive  explanation of a  physical meaning
of the integer parameter $N$ in the radial stochastic process scenario.
One  hypothesis comes from the random-matrix theory, \cite{shukla}, where
$\beta = N-1 = 1, 2, 4$
would correspond to  a number of components of a typical matrix entry which is
decided by the underlying symmetry of the problem (GOE, GUE, GSE). That can be extended
to the case of $N=4$, but there is   no
obvious explanation of that sort  for $N>5$. This issue needs further investigation.

Previously we have indicated  that a common mathematical basis for various level
repulsion mechanisms appropriate to quantum chaos is set by the Calogero-Moser Hamiltonian,
\cite{shukla}.   At the first glance, our stochastic arguments may leave an impression that
something completely divorced from that setting has been obtained in the  present
paper. However things look otherwise  and our theoretical framework proves to
be compatible with standard techniques for spectral analysis of complex quantum systems.

It is peculiar to the general arguments  of Refs. \cite{zambrini,garb} that  invariant
probability
densities give rise to measure preserving stochastic processes in a fully controlled way.
One of  basic ingredients of the  formalism is a proper choice  of  Feynman-Kac
kernel functions, which are the building block for the construction of
 transition probability densities of the pertinent Markov processes.
  Feynman-Kac semigroup operators (and their kernels) explicitly  involve
one particle Hamiltonian operators as generators (in less technical terms
one may think at this point about  rather  standard transformation from the
Fokker-Planck operator to the associated self-adjoint one, \cite{risken}).

For stationary processes, a  general formula relating forward drifts $b(x)$
of the stochastic process
with potentials of the conservative Hamiltonian system reads (we choose a diffusion
coefficient to be equal $1\over 2$), \cite{blanch,garb}:
\be
V(x) = {1\over 2} (b^2 + \nabla \cdot b)\, .
\ee

Upon substituting $b(x)$ according to Eq. (6) we arrive at:
\be
V(x) = {1\over 2} [ {{\beta (\beta -2)}\over {4 x^2}} - (\beta +1) + x^2]
\ee
where $\beta = N-1$. This potential function enters a standard definition of the one
particle Hamiltonian operator (physical parameters have been scaled away):
\be
H = -{1\over 2} \triangle   + V(x)
\ee
where $\triangle = {{d^2}\over {dx^2}}$.  The operator (10) with $V(x)$ defined by (9)
is an equivalent form of  a two-particle  (actually two-level) version of the
Calogero-Moser Hamiltonian, cf. \cite{calogero}
and compare  e.g. the formula (1) in Ref. \cite{shukla}.

Indeed, the classic Calogero-type problem is defined by
\be
H = - {1\over 2}{d^2\over {dx^2}} + {1\over 2}x^2 + {{\beta (\beta - 2)} \over {8x^2}}
\ee
with the well known spectral solution. The eigenvalues read $E_n(\beta )
= 2n + 1 + {1\over 2}[1+\beta (\beta - 2)]^{1/2}$, where  $n\geq 0$ and $\beta >
-1$.

By inspection we can check that all previously considered $N= 2,3,4,5$ radial processes
correspond to the  Calogero  operator of the form  $H - E_0$ where $E_0$
is the ground state  (n=0) eigenvalue. Its explicit from relies on the choice of $\beta $
and   by substituting $\beta = 1,2,3,4$ we  easily check that
$E_0(\beta ) = 1+ {1\over 2}[1+\beta (\beta - 2)]^{1/2}= {1\over 2}(\beta + 1)$
as should be to conform with Eq. (9).

Accordingly, all  considered radial processes arise as  the so-called ground
state processes associated with the Calogero Hamiltonians. Let us recall  that
the classic Ornstein - Uhlenbeck process can be regarded as  the  ground
state process of the harmonic oscillator Hamiltonian operator.
That by the way corresponds to
choosing $N=1$ i.e. $\beta  = 0$ in the above.

It is useful to mention  that our discussion can be readily extended beyond the quantum
chaos context and then to random processes running
in a real time. Incidentally,
Bessel processes with the parameter $\beta $ being not necessarily integer were exploited in
the nonequilibrium statistical formulation for grain growth, \cite{wang,bhakta}.
The apparently unlimited growth can be tamed. By
introducing the harmonic confinement (a substitute for various surface tension effects)
we can produce a family of   grain   growth models with calibrated finite mean
grain size approached  in the asymptotic limit.
In the present problem, we have involved a random walk over  grain sizes  which can
randomly  increase or decrease (that remains in close parallel
to the previously discussed  random walk over erratically   varying level spacings),
 with obvious ergodicity connotations. Indeed,  we may literally  think
about a history and statistics of sizes for a \it single  \rm
  grain evolution in the course of  the growth process (along a single sample trajectory
  in the space of sizes),
or equivalently  about an \it ensemble  \rm statistics, appropriate in
 the large time asymptotic.

{\bf Acknowledgement:} I would like to thank Karol \.{Z}yczkowski  for useful comments on
relationships between small and large random-matrix size spectral predictions.

\end{document}